\documentstyle[preprint,revtex]{aps}
\draft
\begin{document}
\begin{title}
Atomic parity nonconservation and neutron radii in cesium isotopes
\end{title}
\author {B.~Q.~Chen}
\begin{instit}
W.~K.~Kellogg Radiation Laboratory, 106--38\\
California Institute of Technology, Pasadena, CA  91125
\end{instit}
\author{P.~Vogel}
\begin{instit}
Norman Bridge Laboratory of Physics, 161--33\\
California Institute of Technology, Pasadena, CA 91125
\end{instit}
\begin{abstract}
The interpretation of future precise experiments on atomic parity
violation in terms of parameters of the Standard Model could be
hampered by uncertainties in the atomic and nuclear structure. While
the former can be  overcome by measurement in a series of isotopes,
the nuclear structure requires knowledge of the  neutron density. We
use the  nuclear Hartree--Fock method, which includes deformation
effects,  to calculate the proton and neutron densities in
$^{125}$Cs -- $^{139}$Cs. We argue that the good agreement  with the
experimental charge radii, binding energies, and ground  state spins
signifies that the phenomenological nuclear force  and the method of
calculation that we use is adequate.  Based on this agreement, and on
calculations involving different effective interactions,  we estimate
the uncertainties in the differences of the neutron radii  $\delta
\langle r^2 \rangle_{N,N'}$  and conclude that they cause
uncertainties  in the ratio of weak charges, the quantities
determined in the  atomic parity nonconservation experiments, of
less than $10^{-3}$. Such an uncertainty is smaller than the
anticipated experimental error.
\end{abstract}
\section{INTRODUCTION}
Precision studies of electroweak phenomena provide very important
tests of the $SU(2)_L \times U(1)$ Standard Electroweak Model. The
measurement of the parity nonconserving (PNC) components of the
atomic transitions belongs to this class. It offers a unique
opportunity  for testing the electroweak  radiative corrections at
the one loop level, and, possibly, to search for new physics beyond
the standard model  \cite{Lan,Mar}.

The PNC effects in atoms are caused by the $\gamma, Z^0$
interference in the electron--nucleus interaction. The dominant
contribution  comes from the coupling of the axial electronic current
to the vector nuclear current. (The interaction of the electronic
vector current with the nuclear axial current is weaker in heavy
atoms, and can be eliminated by summing over the PNC effects in the
resolved hyperfine components of the atomic transitions. The
hyperfine dependent effect,  which also includes the nuclear anapole
moment, is of interest in its own right \cite{Khr,Hax}, but is not
considered hereafter.) Since the vector current is conserved, atomic
PNC essentially measures the electroweak coupling of the elementary
quarks.

At the present time, PNC measurements in stable $^{133}$Cs atoms have
$\pm$2\% experimental uncertainty \cite{Wie1}. (An earlier experiment
in Cs was performed by Bouchiat  {\it et al.} \cite{Bouch}; the
studies of PNC effects in atoms have been reviewed by Commins
\cite{Com} and Telegdi \cite{Tel}.)  However, improvement by an order
of magnitude in the experimental accuracy is anticipated  and a
possibility of measuring PNC effects in unstable cesium and francium
isotopes has been discussed \cite{Wie2}. At this level, two issues
must be resolved before an interpretation of the PNC data in terms of
the fundamental electroweak couplings  is possible. The atomic
theory, even in its presently most sophisticated form
\cite{Sap,Flam}, introduces about $\pm$1\%  uncertainty. Moreover,
the small but non--negligible effects of nuclear size \cite{For,Poll}
must be addressed. This latter problem is the main topic of the
present work.

Atomic PNC is governed by the  effective bound electron--nucleus
interaction (when taking only the part that remains after averaging
over the hyperfine components) of the form
\begin{equation}
H_{PNC} = \frac{G_F}{2\sqrt{2}} \int [ -N\rho_n ({\bf r })
+ Z (1 - 4 \sin^2 \theta_W ) \rho_p ({\bf r }) ] \times
\psi_e^{\dagger} \gamma_5 \psi_e d^3r ~,
\end{equation}
where the proton and neutron densities $\rho_{p,n}({\bf r })$ are
normalized to unity, and we have assumed the Standard Model nucleon
couplings
\begin{equation}
C_{1p} \equiv 2C_{1u} + C_{1d}
= \frac{1}{2} (1 - 4 \sin^2 \theta_W ) ~,
\end{equation}
\begin{equation}
C_{1n} \equiv C_{1d} + 2C_{1u} = - \frac{1}{2} ~.
\end{equation}
The electron part in Eq.~(1)  can be parametrized as \cite{For,Poll}
\begin{equation}
\rho_5 (r) \equiv
\psi_p^{\dagger} \gamma_5 \psi_s  = C(Z){\cal N}(Z,R)f(r) ~,
\end{equation}
where $C(Z)$ contains all atomic structure effects for a point
nucleus, ${\cal N}$ is a precisely calculable normalization  factor,
and $f(r)$ describes the spatial variation  (normalized such that
$f(0) = 1$).  It is the integrals
\begin{equation}
q_{n,p} = \int f(r) \rho_{p,n}({\bf r })d^3r \label{qf}
\end{equation}
that determine the effect of the proton and neutron distributions  on
the PNC observables.

The formfactors $f(r)$ can be calculated to the order $(Z\alpha)^2$
for a sharp nuclear surface of radius $R$ \cite{For,Poll},
\begin{equation}
f(r) \simeq 1 - \frac{1}{2} (Z\alpha)^2 [(r/R)^2 -
\frac{1}{5}(r/R)^4 ] ~. \label{ff}
\end{equation}
For a diffuse nuclear surface numerical evaluation of $f(r)$ is
necessary (see below). However,  the coefficients at $\langle r^2
\rangle$ and  $\langle r^4 \rangle$ remain numerically of the order
$(Z\alpha)^2$ and depend only weakly on the exact shape of
$\rho_{p,n}({\bf r})$. In addition, since the electric potential near
the nucleus is very strong, one can safely neglect atomic binding
energies in the evaluation of $f(r)$. Below we will separate the
effects of the finite nuclear size (i.e., effects related to the
deviations of $q_{n,p}$ from unity); these terms will be represented
by a nuclear structure correction to the weak charge.

Taking the matrix element of $H_{PNC}$, one obtains
\begin{equation}
\langle i| H_{PNC} |j \rangle
= \frac{G_F}{2\sqrt{2}}C(Z){\cal N}[ Q_W(N,Z) + Q_W^{nuc} (N,Z)] ~,
\end{equation}
where $Q_W(N,Z)$, the quantity of primary interest from the point of
view of testing the Standard Model, is the ``weak charge.''  In the
Standard Model, with couplings (2) and (3), the weak charge is
\begin{equation}
Q_W = -N + Z(1 - 4\sin^2\theta_W) ~. \label{Qweak}
\end{equation}
The nuclear structure correction $Q_W^{nuc}(N,Z)$ describes the part
of the PNC effect that is caused by the finite nuclear size. In the
same approximation as Eq.~(8) above
\begin{equation}
Q_W^{nuc} = -N(q_n - 1) + Z(1 - 4\sin^2\theta_W)(q_p - 1) ~,
\label{Qnuc}
\end{equation}
where $q_{n,p}$ are the integrals of $f(r)$ defined above. (Nuclear
structure also affects the normalization factor ${\cal N}$, which is,
however, determined by the known nuclear charge distribution
\cite{For,Poll}.)

In a measurement that involves several isotopes of the same element,
ratios of the PNC effects depend essentially only on the ratio of the
weak charges and the corresponding nuclear--structure  corrections
$Q_W(N,Z) + Q_W^{nuc}(N,Z)$.  (The dependence  ${\cal N}$ on the
neutron number $N$ will not be considered here.) The ratios of the
nuclear--structure corrected weak charges, in turn, depend,  to a
good approximation, only on the {\it differences} $\Delta q_n$ of the
neutron distributions in the corresponding isotopes. The
uncertainties in these quantities, or equivalently, in the
differences of the neutron mean square radii  $\delta(\Delta \langle
r^2 \rangle_{N,N'})$, then  ultimately limit the accuracy with which
the fundamental parameters, such as $\sin^2 \theta_W$, can be
determined.

It is the purpose of this work to evaluate quantities $q_{n,p}$  for
a number of cesium isotopes, which might be used in future
high--precision PNC experiments \cite{Wie2}.  Moreover, we estimate
the uncertainty in these quantities, respectively in their
differences, since they represent the ultimate limitations for the
interpretation of the PNC measurements.

In section II we describe the nuclear Hartree--Fock calculations that
we performed. In section III we compare the calculated  binding
energies, ground state spins and charge radii with  the experiment.
There we also discuss how corrections for the zero--point vibrational
motion can be estimated and added. From the spread between the
results obtained with two different successful effective Skyrme
forces, and from the pattern of deviations between the calculated and
measured isotope shifts in the charge radii, we then estimate the
uncertainties in the corresponding differences of the neutron radii.
Finally, in section IV, we calculate the  nuclear--structure
corrections to the weak charges $Q_W^{nuc}(Z=55,N=72-84)$ and their
uncertainties and discuss the corresponding limiting uncertainties in
the determination of the fundamental parameters of the Standard
Model. (Our notation follows that of Ref.~\cite{Poll}. Others, e.g.,
Ref.~\cite{Sap} do not explicitly separate the nuclear structure
dependent effects. We believe that such a separation is very useful,
since, as stated above, $f(r)$ in Eq.~(\ref{ff}) and hence also
$q_{n,p}$, Eq.~(\ref{qf}), are essentially  independent of atomic
structure.)

\section{NUCLEAR HARTREE--FOCK CALCULATION}
As demonstrated by numerous calculations, the microscopic description
of nuclear ground state properties by means of the Hartree--Fock (HF)
method with an effective Skyrme force--like interaction  is
remarkably successful \cite{QF78,BF85}. The few adjustable parameters
in the Skyrme force are  chosen to fit the various bulk properties
(energy per  nucleon, compressibility modulus, symmetry energy,
etc.), and properties of several  doubly magic nuclei (binding
energies, charge radii,  etc.) \cite{VB72}. The two most popular sets
of Skyrme parameters, namely Skyrme III  and SkyrmeM$^*$  have been
successfully employed to describe the properties of nuclei in several
regions of the periodic table \cite{BF75,BQ82}. Below we show only a
few formulae essential to the basic understanding of the numerical
calculation that we performed; details can be found in the quoted
references.

The generalized Skyrme force (including all possible spin--exchange
terms and zero--range density--dependent interaction)  can be written
as,
\begin{eqnarray}
V_s = & & t_0 ( 1 + x_0 {P_\sigma}) \delta + {1\over2} t_1 ( 1 + x_1
{P_\sigma})
({\hbox{\font\twelvebi=cmmib10 at 12pt \twelvebi k}}^2 \delta +
\delta {{\hbox{\font\twelvebi=cmmib10 at 12pt \twelvebi k}}}'^2) +
t_2 (1 + x_2) {P_\sigma} {{\hbox{\font\twelvebi=cmmib10 at 12pt
\twelvebi k}}} \cdot
 \delta {{\hbox{\font\twelvebi=cmmib10 at 12pt \twelvebi k}}}'
\nonumber\\
& & + { 1 \over 6} t_3 \rho^{\alpha} \delta +
i W({\hbox{\font\twelvebi=cmmib10 at 12pt \twelvebi \char 27}}_1 +
{\hbox{\font\twelvebi=cmmib10 at 12pt \twelvebi \char 27}}_2) \cdot
{{\hbox{\font\twelvebi=cmmib10 at 12pt \twelvebi k}}} \times \delta
{{\hbox{\font\twelvebi=cmmib10 at 12pt \twelvebi k}}}' ~, \label{SF}
\end{eqnarray}
where $t_{0-3}, x_{0-2}$ and $W$ are the adjustable parameters, and
$\delta \equiv \delta({\bf r - r'})$.

Because we are dealing with odd--A nuclei, the unpaired nucleon
introduces terms that break time--reversal symmetry in  the HF
functional. When the spin degrees of freedom are taken into account,
the breaking of time reversal symmetry leads to a rather complicated
functional \cite{EB75,BF87}. The total energy $E$, which is minimized
in the HF method,  can be written as a space integral of a local
energy density
\begin{equation}
E = \int {\cal H}({{\hbox{\font\twelvebi=cmmib10 at 12pt \twelvebi
r}}}) d^3 {{\hbox{\font\twelvebi=cmmib10 at 12pt \twelvebi r}}} ~,
\end{equation}
with
\begin{eqnarray}
{\cal H}({\bf r}) = & &{\hbar^2 \over 2 m } \tau +
B_1 \rho^2 + B_2 (\rho_n^2 + \rho_p^2) + B_3 (\rho \tau -
{\hbox{\font\twelvebi=cmmib10 at 12pt \twelvebi j}}^2) +
B_4 ( \rho_n \tau_n - {\hbox{\font\twelvebi=cmmib10 at 12pt \twelvebi
j}}_n^2 + \rho_p \tau_p - {\hbox{\font\twelvebi=cmmib10 at 12pt
\twelvebi j}}_p^2)
\nonumber \\
& & + B_5 \rho \Delta \rho + B_6 (\rho_n \Delta \rho_n + \rho_p
\Delta \rho_p) +
B_7 \rho^{2+\alpha} + B_8 \rho^{\alpha} (\rho_n^2 + \rho_p^2)
\nonumber \\
& & + B_9 (\rho {\hbox{\font\twelvebs=cmbsy10 at 12pt \twelvebs\char
"072}} \cdot {\hbox{\font\twelvebi=cmmib10 at 12pt \twelvebi J}} +
{\hbox{\font\twelvebi=cmmib10 at 12pt \twelvebi j}} \cdot
{\hbox{\font\twelvebs=cmbsy10 at 12pt \twelvebs \char "072}} \times
{\hbox{\font\twelvebi=cmmib10 at 12pt \twelvebi \char 26}} + \rho_n
{\hbox{\font\twelvebs=cmbsy10 at 12pt \twelvebs \char "072}} \cdot
{\hbox{\font\twelvebi=cmmib10 at 12pt \twelvebi J}}_n +
{\hbox{\font\twelvebi=cmmib10 at 12pt \twelvebi j}}_n \cdot
{\hbox{\font\twelvebs=cmbsy10 at 12pt \twelvebs \char "072}}
\times {\hbox{\font\twelvebi=cmmib10 at 12pt \twelvebi \char 26}}_n
+ \rho_p {\hbox{\font\twelvebs=cmbsy10 at 12pt \twelvebs \char "072}}
\cdot {\hbox{\font\twelvebi=cmmib10 at 12pt \twelvebi J}}_p +
{\hbox{\font\twelvebi=cmmib10 at 12pt \twelvebi j}}_p \cdot
{\hbox{\font\twelvebs=cmbsy10 at 12pt \twelvebs \char "072}} \times
{\hbox{\font\twelvebi=cmmib10 at 12pt \twelvebi \char 26}}_p)
 \nonumber \\
& & + B_{10} {\hbox{\font\twelvebi=cmmib10 at 12pt \twelvebi \char
26}}^2 + B_{11} ({\hbox{\font\twelvebi=cmmib10 at 12pt \twelvebi
\char 26}}_n^2 + {\hbox{\font\twelvebi=cmmib10 at 12pt \twelvebi
\char 26}}_p^2) + B_{12}
 \rho^{\alpha}
{\hbox{\font\twelvebi=cmmib10 at 12pt \twelvebi \char 26}}^2 + B_{13}
\rho^{\alpha} ({\hbox{\font\twelvebi=cmmib10 at 12pt \twelvebi \char
26}}_n^2 + {\hbox{\font\twelvebi=cmmib10 at 12pt \twelvebi \char
26}}_p^2) +
E_C ~. \label{HFD}
\end{eqnarray}
For complete expressions of the Coulomb energy $E_C$ and the
coefficients  $B_i ( i = 1, \dots , 13 )$ see Ref.~\cite{BF87}, where
the dependence on Skyrme force parameters in Eq.~(\ref{SF}) is given.
The mass densities $\rho_\tau$, kinetic density $\tau_\tau$,  current
density ${\hbox{\font\twelvebi=cmmib10 at 12pt \twelvebi j}}_\tau$,
spin--orbit density ${\hbox{\font\twelvebs=cmbsy10 at 12pt \twelvebs
\char "072}} \cdot {\hbox{\font\twelvebi=cmmib10 at 12pt \twelvebi
J}}_\tau$ and vector density ${\hbox{\font\twelvebi=cmmib10 at 12pt
\twelvebi \char 26}}_\tau (\tau = {n, p})$ in Eq.~(\ref{HFD}) can, in
turn, be expressed in terms of the single--particle wave functions
$\Phi_k$.  The variation of $E$ with respect to
$\Phi_k^*({\hbox{\font\twelvebi=cmmib10 at 12pt \twelvebi r}},
\sigma)$ defines the one--body Hartree--Fock hamiltonian $h$
\cite{BF87}.

In the following we will use the mass densities $\rho_\tau$, which
can be expressed as
\begin{equation}
\rho({\hbox{\font\twelvebi=cmmib10 at 12pt \twelvebi r}})  =
\sum_{k,\sigma} v_k^2 | \Phi_k({\hbox{\font\twelvebi=cmmib10 at 12pt
\twelvebi r}}, \sigma)|^2 ~.
\end{equation}
Here $\Phi_k({\hbox{\font\twelvebi=cmmib10 at 12pt \twelvebi r}},
\sigma)$ denotes the component of the $k$th  single--nucleon wave
function with spin ${1\over2} \sigma (\sigma = \pm 1)$ along the $z$
direction, and $v_k^2$ are the BCS occupation  factors (see below).
The expressions for the other densities are again given in
Ref.~\cite{BF87}.

The mean square proton and neutron radii are given by the usual
formulae
\begin{equation}
r^2_\tau = \int r^2 \rho_\tau({\hbox{\font\twelvebi=cmmib10 at 12pt
\twelvebi r}}) d^3 {\hbox{\font\twelvebi=cmmib10 at 12pt \twelvebi
r}} ~.
\end{equation}

In this work, two discrete symmetries, namely parity and
$z$--signature, are imposed on the wave functions \cite{BF85,BF87}.
The complete description of a wave function requires four real
functions corresponding to the real, imaginary, spin--up and
spin--down parts of $\Phi_k$ \cite{BF87}.

The numerical approximation to the HF energy $E$ is obtained by a
discretization of the configuration space on a three--dimensional
rectangular mesh. The mesh size $\Delta x$ is the same in the three
directions and the abscissae of the mesh points are  ${1\over2} (2 n
+ 1) \Delta x$. In this work, $\Delta x$ is  $0.8~\hbox{fm}$, and the
mesh size is  $16 \times 16 \times 16$. The numerical procedure is
described in detail in Ref.~\cite{BF85}.

Pairing correlations need to be included in a realistic description
of  medium and heavy nuclei. We choose to describe pairing between
identical nucleons within the BCS formalism using a  constant
strength seniority force \cite{BF85}. In the usual BCS scheme, the
paired states are assumed to be the two time--reversed orbitals
$\Phi_k$ and $\Phi_{\hat k}$. Although time reversal symmetry is
broken in our calculations of odd--A nuclei, the time--reversal
breaking terms in the functional generated by the unpaired odd
nucleon are very small compared to the time--reversal conserving
terms so that the time reversal symmetry  is still approximately
good. In our calculation we define the pairing partner $\Phi_{\hat
k}$  of state $\Phi_k$ to be the eigenstate of $h$ whose overlap with
$\hat T \Phi_k$ is maximal ($\hat T$ is the time--reversal
operator). Because the single particle orbital occupied by the
unpaired nucleon and its signature partner do not contribute to the
pairing energy, we introduce blocking in our code to prevent these
two orbitals from participating in  pairing and force their BCS
occupation numbers to be $1$ and $0$, respectively.

As some of the cesium isotopes considered here are deformed, it is
very important to take the deformation degrees of freedom into
account. The method of solving the HF+BCS equations by discretization
of the wave functions on a rectangular mesh allows any type of even
multipole deformation. The deformation energy curves are obtained by
a constraint on the mass quadrupole tensor $Q_{ij} = (3 x_i x_j - r^2
\delta_{ij})$. The two discrete symmetries of the wave functions
$\Phi_k$ ensure that the principal axes of inertia lie along the
coordinate axes.  The quadrupole tensor is, therefore, diagonal and
its principal values $Q_i$ can be expressed in terms of two
quantities $Q_0$ and $\gamma$ as
\begin{equation}
Q_i = Q_0 \cos(\gamma + i {2 \over 3} \pi ), \quad i = 1, 2, 3 ~,
\end{equation}
where $Q_0$ and $\gamma$ satisfy the inequalities
\begin{equation}
Q_0 \ge 0, \quad 0 \le \gamma \le {2 \over 3} \pi ~.
\end{equation}
The values of the three constraints $Q_i$ were computed from the
desired values of $Q_0$ and $\gamma$ and inserted in a quadratic
constraint functional added to the variational  energy, according to
the method described in Ref.~\cite{FQ73}. In the calculations
described below, we constrain the nuclear  shape to be axially
symmetric ($\gamma = 0$).

\section{COMPARISON WITH EXPERIMENT}

In Fig.~1 we show the potential energy curves for $^{125}$Cs --
$^{139}$Cs.  According to our calculations  with SkyrmeIII (SkmIII)
and SkyrmeM* (SkM*) forces the lighter cesium isotopes $N \leq 76$
are deformed. For SkmIII such an assignment is able to explain the
observed ground state spins of  $\frac{1}{2}^+$ for $N = 70 - 74$ and
$\frac{5}{2}^+$ for $N = 76$.  For SkM* the mean field proton states
$g_{7/2}$ and $d_{5/2}$ are interchanged and therefore the ground
state spin assignments for the deformed cesium isotopes are not
correct. (This turns out not to be a very crucial problem.) Binding
energies and shifts $\delta r_{p,n}^2$ and $\delta r_{p,n}^4$
calculated with the SkM* and SkmIII interactions are shown in Tables
I and II. The binding energies agree in both cases with the
experimental values with largest deviation of 4~MeV  out of about
1000~MeV of total binding energy.

The comparison between the measured and calculated isotope shifts is
illustrated in Figs.~2 and 3 as a series of successively better
approximations. First, the crosses, connected by dashed lines to
guide eyes, show the isotope shifts for spherical nuclei. The
agreement with experiment is not very good even though the spherical
calculation correctly predicts that the slope of the dependence
$\delta r_p^2(A)$ is about half of the slope expected from the simple
relation $R = r_0 A^{1/3}$. This means that, on average, the
neutron--proton interaction we use has the correct magnitude.

Next, the equilibrium deformation for the lighter cesium isotopes is
included (open squares), leading to a much better agreement. Further
improvement is achieved when the effect of zero--point quadrupole
vibrational motion is taken into account.  It is well known that the
mean square radius of a vibrating nucleus is increased by \cite{BMI}
\begin{equation}
\langle r^2 \rangle_{\beta} = \langle r^2 \rangle_0 ( 1 +
\frac{5}{4\pi} \langle \beta^2 \rangle ) ~. \label{rb}
\end{equation}
We include this effect of the shape fluctuations using the quantities
$\langle \beta^2 \rangle$ extracted from the measured transition
matrix elements  $B(E2,0^+ \rightarrow 2^+)$ and the relation
\begin{equation}
\langle \beta^2 \rangle = B(E2,0^+ \rightarrow 2^+)
[3 Z R_0^2 / 4 \pi ]^{-2} ~.
\end{equation}
We take the average $B(E2)$ of the corresponding Xe and Ba isotopes
with neutron numbers $N=78-84$ and correct the radii of
$^{133}$Cs--$^{139}$Cs accordingly, as shown in Figs.~2 and 3. Thus,
further improvement in the comparison with the measured isotope
shifts results.  (For $N=84$ the $B(E2)$ values are not known. We use
instead the empirical relation between the energy of the lowest $2^+$
state and the deformation parameter $B(E2)$ \cite{Ram}.) This
correction results in changes in $r^2$ of 0.2124~fm$^2$ in
$^{133}$Cs, 0.1325~fm$^2$ in $^{135}$Cs, 0.0724~fm$^2$ in $^{137}$Cs,
and 0.1263~fm$^2$ in $^{139}$Cs.

In a fully consistent calculation, one should make a similar
correction for the deformed cesium  isotopes as well. Since the
corresponding $B(E2)$ values for the vibrational states are not
known, and the corrections are expected to be small, we do not make
them. Instead, we somewhat arbitrarily assume that the zero--point
motion correction is the same as in the semimagic $^{137}$Cs. We
believe this explains the somewhat poorer agreement in the deformed
cesium isotopes.

Even though the quadrupole $2^+$ states contribute most to the mean
square radius via Eq.~(\ref{rb}), other vibrational  states, e.g.,
the octupole $3^-$ and the giant resonances,  contribute as well;
however, all such states  not only have smaller collective amplitudes
but, even more  importantly, vary more smoothly with the atomic mass
(or neutron number) than the $2^+$ states, and hence their
contribution to the shifts $\delta r^2$ should be correspondingly
smaller.

Altogether, the error in the shift $\delta r_p^2$ is at most
0.2~fm$^2$, and appears to be independent of the change in  the
neutron number $\Delta N$. Thus, for the following  considerations we
assign an uncertainty in the relative value of $\delta r_p^2$ of
0.2~fm$^2$. Very little is known experimentally about the moments
$r_p^4$. Quite conservatively, we assume that the uncertainty in
$\delta r_p^4$ is  $\langle r_p^2 \rangle \times \Delta r_p^2 \simeq$
5~fm$^4$.

Before turning our attention to the neutron radii, it is  worthwhile
to make a brief comment about the comparison with $absolute$ values
of $\langle r_p^2 \rangle$ and $\langle r_p^4 \rangle$.
Experimentally,  muonic x--ray energies for the stable $^{133}$Cs
have been fitted to the Fermi distribution with the halfway radius
$c$ = 5.85~fm, surface thickness $t$ = 1.82~fm \cite{Eng,Wu}, and
$\langle r_p^2 \rangle$ = 23.04~fm$^2$.  Such a Fermi distribution
gives $\langle r_p^4 \rangle$ = 673~fm$^4$.  Our HF calculation
corrected for zero--point vibrational motion with $\langle \beta^2
\rangle$ = 0.024, as described above, gives $\langle r_p^2
\rangle_{HF}$ = 23.27~fm$^2$ for SkmIII and 22.69~fm$^2$ for SkM*
interaction, both quite close to the experimental value. The
calculated $\langle r_p^4 \rangle$ moments (not corrected for the
zero--point motion) are 671(SkmIII) and 652(SkM*)~fm$^4$.  We see,
therefore, that the calculation is quite successful in the absolute
radii (and even surface thicknesses), in particular for the SkmIII
interaction (which gives also the correct ground state spin).

The calculated shifts in the neutron radii $\delta r_n^2$ and $\delta
r_n^2$ are listed in Tables I (SkM*)  and II (SkmIII) and the
quantities $\delta r_n^2$ corrected for the effect of zero--point
vibrational motion are displayed in Fig.~4. Several comments about
these are in order. First, the slope of the dependence of $\delta
r_n^2 (A)$ for spherical configurations is correspondingly steeper
than the slope following from $R = r_0 A^{1/3}$. That is obviously a
correct  result; the combination of a smaller slope in the proton
radii and a larger slope in the neutron radii when neutrons are added
is necessary to maintain on average the $R = r_0 A^{1/3}$ relation.
Second, the HF calculations imply that the proton and neutron
distributions have essentially identical deformations. This agrees
with the general conclusion about the isoscalar character of
low--frequency collective modes in nuclei (see, e.g.,
Ref.~\cite{BMII}). Thus, we accept this result and do not assign any
additional uncertainty to the possible difference in the deformation
of protons and neutrons. Finally, for the same reason, we use the
same $B(E2)$ values, and the $\langle \beta^2 \rangle$ extracted from
them, to correct the neutron radii using Eq.~(\ref{rb}). Assuming all
of the above, we assign {\it identical} uncertainties to the neutron
shifts $\delta r_n^2$ and the proton shifts $\delta r_p^2$,  and
similarly to the fourth moments $\delta r_{n,p}^4$.

Very little reliable experimental information on neutron distribution
in nuclei is available. In Ref.~\cite{GNO}, data from pionic atoms
are analyzed. The corresponding best fit for neutron mean square
radii agrees very well with the HF results quoted there. The nearest
nucleus to cesium in Ref.~\cite{GNO} is $^{142}$Ce. Scaling it with
$A^{2/3}$ one arrives at $\langle r_n^2 \rangle$ = 24.7~fm$^2$ for
$^{133}$Cs, somewhat larger than our calculated values 23.7 and 24.0
for SkM* and SkmIII, respectively. In Ref.~\cite{Sap}, the
theoretical neutron density of Brack {\it et al.} \cite{Brack} with
$\langle r_n^2 \rangle$ = 23.5~fm$^2$, was used. That value,
presumably obtained by interpolation from the values obtained by the
HF method using the SkM* interaction, is, not surprisingly, quite
close to our calculated values. This limited comparison suggests that
the absolute radii $\langle r_n^2 \rangle$  have  uncertainties of
about 1~fm$^2$. The uncertainty in the shifts $\delta r_n^2$ should
be substantially smaller, and our estimated error of 0.2~fm$^2$ does
not seem unreasonable.

In Ref.~\cite{Poll} the uncertainty in the integrals $q_{n,p}$ was
estimated from the spread of the calculated values with a wide
variety of interactions. Some of the interactions employed in
\cite{Poll} give better agreement for known quantities (charge radii,
binding energies, etc.) than others. We chose to use only the two
most successful interactions. The spread in the calculated shifts
$\delta r_{p,n}^2$ for these two interactions is less than our
postulated error of 0.2~fm$^2$.

Pollock {\it et al.} \cite{Poll} also argue that the isovector
surface term  $(\rho_p - \rho_n) \nabla^2 (\rho_p - \rho_n)$ in the
Skyrme Lagrangian is poorly determined and may affect the neutron
skin significantly, without affecting most bulk nuclear properties.
We tested this claim by modifying  simultaneously the coefficients
$B_5 \rightarrow B_5(1+x)$ and $B_6 \rightarrow B_6-2B_5x$ in
Eq.~(\ref{HFD}). We find that when we vary  $x$ (i.e., the relative
strength of the isovector surface term) from +0.3 to -0.3 the proton
radius $\langle r_p^2 \rangle$ changes indeed very little (about
0.06~fm$^2$) and the neutron radius changes somewhat more (by about
0.1~fm$^2$, still less than our estimated error). However, the
binding energy changes by about 5~MeV, more than the largest
discrepancy between the theory and experiment. Thus, we do not think
that the uncertainty in this particular coefficient of the Skyrme
force  alters our conclusions.

\section{ESTIMATED UNCERTAINTIES IN PNC EFFECTS}

The nuclear structure effects are governed by the coefficients
$q_{n,p}$, Eq.~(\ref{qf}), which in turn involve integrals of the
formfactors $f(r)$, Eq.~(\ref{ff}). The function $f(r)$ is slowly
varying over the nuclear volume, and may be accurately approximated
by a power series
\begin{equation}
f(r) = 1 + f_2 \times r^2 + f_4 \times r^4 ~,
\end{equation}
and, therefore,
\begin{equation}
q_{n,p} = 1 + f_2 \times \langle r_{n,p}^2 \rangle +
f_4 \times \langle r_{n,p}^4 \rangle ~.
\end{equation}

For a sharp nuclear surface density distribution  the only relevant
parameter is the nuclear radius $R$ and  $\langle r^{2n} \rangle =
3/(2n+3)R^{2n}$. Using the experimental $\langle r^2 \rangle$ =
23.04~fm$^2$ for $^{133}$Cs \cite{Eng},  we find from Eq.~(\ref{ff})
\begin{equation}
f(r) = 1 - 2.10 \times 10^{-3} r^2
+ 1.09 \times 10^{-5} r^4 ~,
\end{equation}
where the distance is measured in fermis. If, instead, we solve
numerically the Dirac equation for the  $s_{1/2}$ and $p_{1/2}$ bound
electron states in the field of the finite size diffuse surface
nucleus, we obtain the coefficients  $f_2 (f_4)$ of
$-2.31\times10^{-3} (1.21\times10^{-5})$ when we use the standard
surface thickness parameter $t=2.25$~fm, and $-2.267\times10^{-3}
(1.157\times10^{-5})$ when we use the surface thickness adjusted so
that the nuclear density parametrized by the two--parameter Fermi
distribution resembles as closely as possible the Hartree--Fock
charge density in $^{133}$Cs.

The expansion coefficients $f_2, f_4$ depend, primarily, on the mean
square charge radius. To take this dependence into account, we use
for $^{133}$Cs the $f_2$ and $f_4$ above, and for the other isotopes,
we use the same surface thickness parameter $(t=1.82)$ and adjust the
halfway radius in such a way that the experimental $<r_p^2>$ are
correctly reproduced.

It is easy now to evaluate the uncertainty in the factors $q_{n,p}$
given the coefficients $f_2, f_4$ and our estimates of the
uncertainties in $\langle r^2 \rangle$ and $\langle r^4 \rangle$.
Substituting the corresponding values, we find that the uncertainty
is $\delta q_{n,p}$ = 4.6$\times$10$^{-4}$, caused almost entirely by
the uncertainty in the mean square radii  $\langle r_{n,p}^2
\rangle$. This uncertainty represents about 1\% of the deviations of
$q_{n,p}$  values from unity.

Before evaluating the nuclear structure corrections $Q_W^{nuc}(N,Z)$
we have to consider the effect of the intrinsic nucleon structure.
Following \cite{Poll} we use
\begin{equation}
q_{p,n}^{int} = \int d^3{\bf r} \frac{1}{6}
\langle r^2 \rangle_{int,(p,n)}^w f(r)
\nabla^2 \rho_{p,n}/Q_{p,n}^w ~,
\end{equation}
where $\langle r^2 \rangle_{int}^w$ are the nucleon weak radii, and
$Q_{p,n}^w$ are the nucleon weak charges. Neglecting the
``strangeness radius'' of the nucleon, and using the fitted
two--parameter Fermi density distribution, we find
\begin{equation}
q_p^{int} = -0.00290, ~~ q_n^{int} = -0.00102 ~,
\end{equation}
very close to the sharp nuclear surface values of Pollock {\it et
al.} \cite{Poll}. The above intrinsic nucleon structure corrections
are small, but not negligible. More importantly, they are independent
of the nuclear structure, and cancel out in the differences $\Delta
q_{n,p}$.

The quantities 100$\times$($q_n$-1) and 100$\times$($q_p$-1) are
listed in Table III for all cesium isotopes and for the two Skyrme
interactions we consider. One can see that they vary by about 4\% for
neutrons and are essentially constant for protons when the neutron
number increases from $N$ = 70 to 84. The variation with $N$ is
essentially identical for the two forces, while the small difference
between the $q_{n,p}$ values calculated with the two forces  reflects
the difference in the {\it absolute} values of radii for the two
interactions.

The weak charges $Q_W(N,Z)$ and the nuclear structure corrections
$Q_W^{nuc}(N,Z)$ in Table III are radiatively corrected. Thus,
instead of the  formulae (\ref{Qweak}), (\ref{Qnuc}) we use
\begin{equation}
Q_W(N,Z) = 0.9857 \times [-N + Z(1 - 4.012\bar{x})] ~,~
\bar{x} = 0.2323 + 0.00365S ~,
\end{equation}
following \cite{Mar}.  Here $S$ is the parameter characterizing the
isospin--conserving ``new'' quantum loop corrections \cite{PT}. Also,
\begin{equation}
Q_W^{nuc}(N,Z) = 0.9857 \times [-N(q_n - 1) +
Z(1 - 4.012\bar{x})(q_p - 1)] ~.
\end{equation}
These quantities, evaluated for $S = 0$, are shown in Table III. The
assumed uncertainty in the shifts of the mean square radii, and
consequently in the  changes in factors $q_{n,p}$ results in the
relative uncertainty $\delta Q_W/Q_W$ of  5$\times$10$^{-4}$. That
uncertainty, therefore, represents the ``ultimate'' nuclear structure
limitation on the tests of the Standard Model in the atomic PNC
experiments involving several isotopes.

In the atomic PNC experiments involving a {\it single} isotope, the
uncertainty in the neutron mean square radius is larger, and 1~fm$^2$
appears to be a reasonable choice. Thus, from nuclear structure
alone, the weak charge in a single isotope has relative uncertainty
of about 2.5$\times$10$^{-3}$, perhaps comparable to the best
envisioned measurements, but considerably smaller then the present
uncertainty associated with the {\it atomic} structure.

Suppose now that in an experiment involving several cesium isotopes
one is able to determine the ratio
\begin{equation}
R(N',N) = \frac{Q_W(N',Z)+Q_W^{nuc}(N',Z)}{Q_W(N,Z)+
Q_W^{nuc}(N,Z)}
\end{equation}
with some relative uncertainty $\delta R/R$. To a (reasonable) first
approximation
\begin{equation}
R(N',N) \approx \frac{Q_W(N',Z)}{Q_W(N,Z)}\times[1
+ q_n(N') - q_n(N)] ~.
\end{equation}
Thus, we see that nuclear structure contributes to the uncertainty of
$R$ at the level of roughly 7$\times$10$^{-4}$, where we added the
individual errors in quadrature. This  uncertainty is much smaller
than the anticipated experimental error.

In such a measurement, therefore, the uncertainty in $\bar{x}$ will
be
\begin{equation}
\frac{\delta \bar{x}}{\bar{x}} \approx \frac{\delta R}{R}\times
\frac{NN'}{Z\Delta N} \approx 8\frac{\delta R}{R}~,
\end{equation}
(see also \cite{Mar,Poll}) where the last factor is evaluated for
$N',N$ = 70, 84. The above equation illustrates the obvious advantage
of using isotopes with large $\Delta N$. Also, by performing the
measurement with several isotope pairs, one can further decrease the
uncertainty  $\delta \bar{x}$. On the other hand, the uncertainty in
the important parameter $S$ is determined from the relation $\delta
\bar{x} = 0.00365\delta S$, and thus
\begin{equation}
\delta S \approx \frac{\delta R}{R}\times
\frac{NN'}{0.014Z\Delta N} ~.
\end{equation}

In conclusion, we have evaluated the nuclear structure  corrections
to the weak charges for a series of cesium isotopes, and estimated
their uncertainties. We concluded that the imperfect knowledge of the
neutron distribution in cesium isotopes does not represent in the
foreseeable future a limitation on the accuracy with which the
Standard Model could be tested in the atomic PNC experiments.
\acknowledgments We would like to thank C.~Wieman and D.~Vieira whose
discussion of the proposed experiments inspired the work described
here. This research was performed in part using the Intel Touchstone
Delta System operated by Caltech on behalf of the Concurrent
Supercomputing Consortium. This work was supported in part by the
U.S. Department of Energy under Contract \#DE--F603--88ER--40397, and
by the National Science Foundation, Grant No.~PHY90--13248.

\figure{
The potential energy curves for the isotopes $^{125}$Cs - $^{139}$Cs
calculated by the Hartree--Fock method using the SkmIII interaction.
The notations are shown on the figure.
\label{def}}

\figure{
Calculated and experimental isotope shifts  $\delta \langle r_p^2
\rangle$ in cesium, normalized to the semimagic $^{137}$Cs. The SkM*
interaction has been used. The correction for zero--point
vibrations is described in the text. The following notations are
used: experimental isotope shift $\Diamond$, spherical HF isotope
shifts $+$, HF including equilibrium deformation $\Box$, and
corrected for zero--point vibrations $\times$. \label{rpstar}}

\figure{
Calculated and experimental isotope shifts  $\delta \langle r_p^2
\rangle$ in cesium. The SkmIII
interaction has been used. The correction for zero--point vibrations
is described in the text. The same notations used in
Fig.~\ref{rpstar} are used here. \label{rpIII}}

\figure{
Calculated changes in the neutron radii $\delta \langle r_n^2
\rangle$ in cesium. The results,
corrected for zero--point vibrational motion, and calculated with the
SkmIII ($\Diamond$) and SkM* ($+$) interactions, are shown.}

\widetext
\begin{table}
\caption{Results of the Hartree--Fock calculations with the SkM*
interactions. The experimental binding energies and
isotope shifts $\delta \langle r_p^2 \rangle$ are also listed
for comparison. (The binding energies are in~MeV, all radial moments
in~fm.) The experimental isotope shifts are from Ref.~\cite{Thi},
normalized to the stable isotope $^{133}$Cs.}
\label{Tab. 1}
\begin{tabular}{cccrrrrrrr}
N & B & B$_{\hbox{\small HF}}$ & \multicolumn{1}{c}{$\delta
r_p^2$(exp)} &
\multicolumn{1}{c}{$\delta r_p^2$} & \multicolumn{1}{c}{$\delta
r_p^2$(sph.)} &
\multicolumn{1}{c}{$\delta r_p^4$} & \multicolumn{1}{c}{$\delta
r_n^2$} &
\multicolumn{1}{c}{$\delta r_n^2$(sph.)} & \multicolumn{1}{c}{$\delta
r_n^4$}\\
\tableline
 70 &  1049.98 &  1045.82 &  -0.1517 &  -0.0899 &  -0.4445 &   7.987
&  -0.6803 &  -1.0787 & -31.126 \\
 72 &  1068.25 &  1064.38 &  -0.0985 &  -0.0348 &  -0.3285 &   8.836
&  -0.4603 &  -0.7931 & -19.563 \\
 74 &  1085.66 &  1082.15 &  -0.0561 &  -0.0199 &  -0.2161 &   6.247
&  -0.2927 &  -0.5186 & -11.931 \\
 76 &  1102.37 &  1099.36 &  -0.0141 &   0.0090 &  -0.1070 &   4.306
&  -0.1253 &  -0.2544 &  -4.538 \\
 78 &  1118.52 &  1117.69 &   0.0000 &   0.0000 &   0.0000 &   0.000
&   0.0000 &   0.0000 &   0.000 \\
 80 &  1134.24 &  1135.71 &   0.0250 &   0.1054 &   0.1054 &   4.872
&   0.2454 &   0.2454 &  14.025 \\
 82 &  1149.27 &  1152.18 &   0.0821 &   0.2531 &   0.2531 &   9.658
&   0.5132 &   0.5132 &  28.754 \\
 84 &  1159.57 &  1164.16 &   0.3604 &   0.3394 &   0.3394 &  17.820
&   0.8866 &   0.8866 &  59.902 \\
\end{tabular}
\end{table}
\begin{table}
\caption{Results of the Hartree--Fock calculations with the SkmIII
interactions. The experimental binding energies and
isotope shifts $\delta \langle r_p^2 \rangle$ are also listed
for comparison. (The binding energies are in~MeV, all radial moments
in~fm.) The experimental isotope shifts are from Ref.~\cite{Thi},
normalized to the stable isotope $^{133}$Cs.}
\label{Tab. 2}
\begin{tabular}{cccrrrrrrr}
N & B & B$_{\hbox{\small HF}}$ & \multicolumn{1}{c}{$\delta
r_p^2$(exp)} &
\multicolumn{1}{c}{$\delta r_p^2$} & \multicolumn{1}{c}{$\delta
r_p^2$(sph.)} &
\multicolumn{1}{c}{$\delta r_p^4$} & \multicolumn{1}{c}{$\delta
r_n^2$} &
\multicolumn{1}{c}{$\delta r_n^2$(sph.)} & \multicolumn{1}{c}{$\delta
r_n^4$}\\
\tableline
 70 &  1049.98 &  1047.12 &  -0.1517 &  -0.1322 &  -0.5097 &   7.670
&  -0.5484 &  -1.0265 & -24.683 \\
 72 &  1068.25 &  1065.52 &  -0.0985 &  -0.1015 &  -0.3813 &   6.023
&  -0.4141 &  -0.7592 & -18.954 \\
 74 &  1085.66 &  1083.44 &  -0.0561 &  -0.0440 &  -0.2536 &   6.317
&  -0.2526 &  -0.4991 & -11.388 \\
 76 &  1102.37 &  1100.62 &  -0.0141 &  -0.0096 &  -0.1265 &   3.117
&  -0.1096 &  -0.2461 &  -5.198 \\
 78 &  1118.52 &  1118.01 &   0.0000 &   0.0000 &   0.0000 &   0.000
&   0.0000 &   0.0000 &   0.000 \\
 80 &  1134.24 &  1134.75 &   0.0250 &   0.1254 &   0.1254 &   6.530
&   0.2392 &   0.2392 &  14.634 \\
 82 &  1149.27 &  1153.20 &   0.0821 &   0.2508 &   0.2508 &  13.124
&   0.4721 &   0.4721 &  29.191 \\
 84 &  1159.57 &  1161.94 &   0.3604 &   0.4120 &   0.4120 &  22.346
&   0.8674 &   0.8674 &  59.984 \\
\end{tabular}
\end{table}
\begin{table}
\caption{The radiatively corrected weak charges $Q_W(N,Z)$,
nuclear structure corrections $Q_W^{nuc}(N,Z)$, and the
quantities $q_n$-1, $q_p$-1 (the factors ($q_{p,n}$-1)
contain the intrinsic nucleon structure correction, and are
multiplied by 100 for easier display) calculated with
the SkM* and SkmIII interactions, and with the vibrational
corrections described in the text.}
\label{Tab. 3}
\begin{tabular}{|cc|ccc|ccc}
\multicolumn{2}{|c|}{}&\multicolumn{3}{c|}{SkM*}
&\multicolumn{3}{c|}{SkmIII}\\
N &$Q_W(N,Z)$&$Q_W^{nuc}(N,Z)$&$q_n$-1&$q_p$-1&$Q_W^{nuc}(N,Z)$
&$q_n$-1 &\multicolumn{1}{c|}{ $q_p$-1}\\
\hline
70&-65.312&2.967&-4.55&-4.64&3.015&-4.62&-4.74~\vline\\
72&-67.283&3.077&-4.58&-4.64&3.118&-4.64&-4.74~\vline\\
74&-69.254&3.184&-4.60&-4.64&3.225&-4.66&-4.75~\vline\\
76&-71.226&3.291&-4.62&-4.64&3.330&-4.68&-4.75~\vline\\
78&-73.197&3.422&-4.68&-4.68&3.458&-4.73&-4.79~\vline\\
80&-75.169&3.528&-4.69&-4.67&3.564&-4.74&-4.79~\vline\\
82&-77.140&3.638&-4.71&-4.68&3.669&-4.76&-4.79~\vline\\
84&-79.112&3.745&-4.73&-4.66&3.780&-4.78&-4.78~\vline\\
\end{tabular}
\end{table}

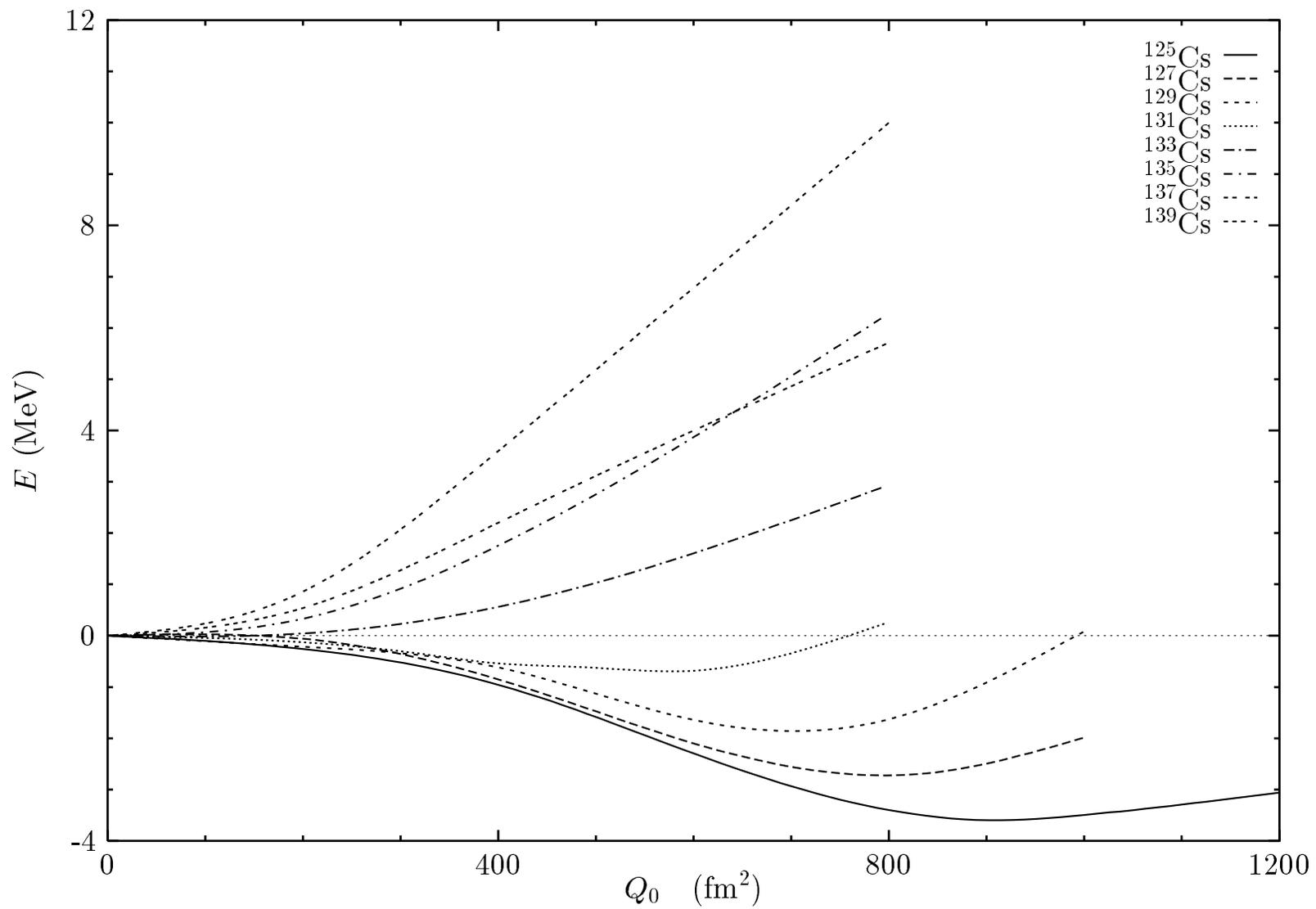
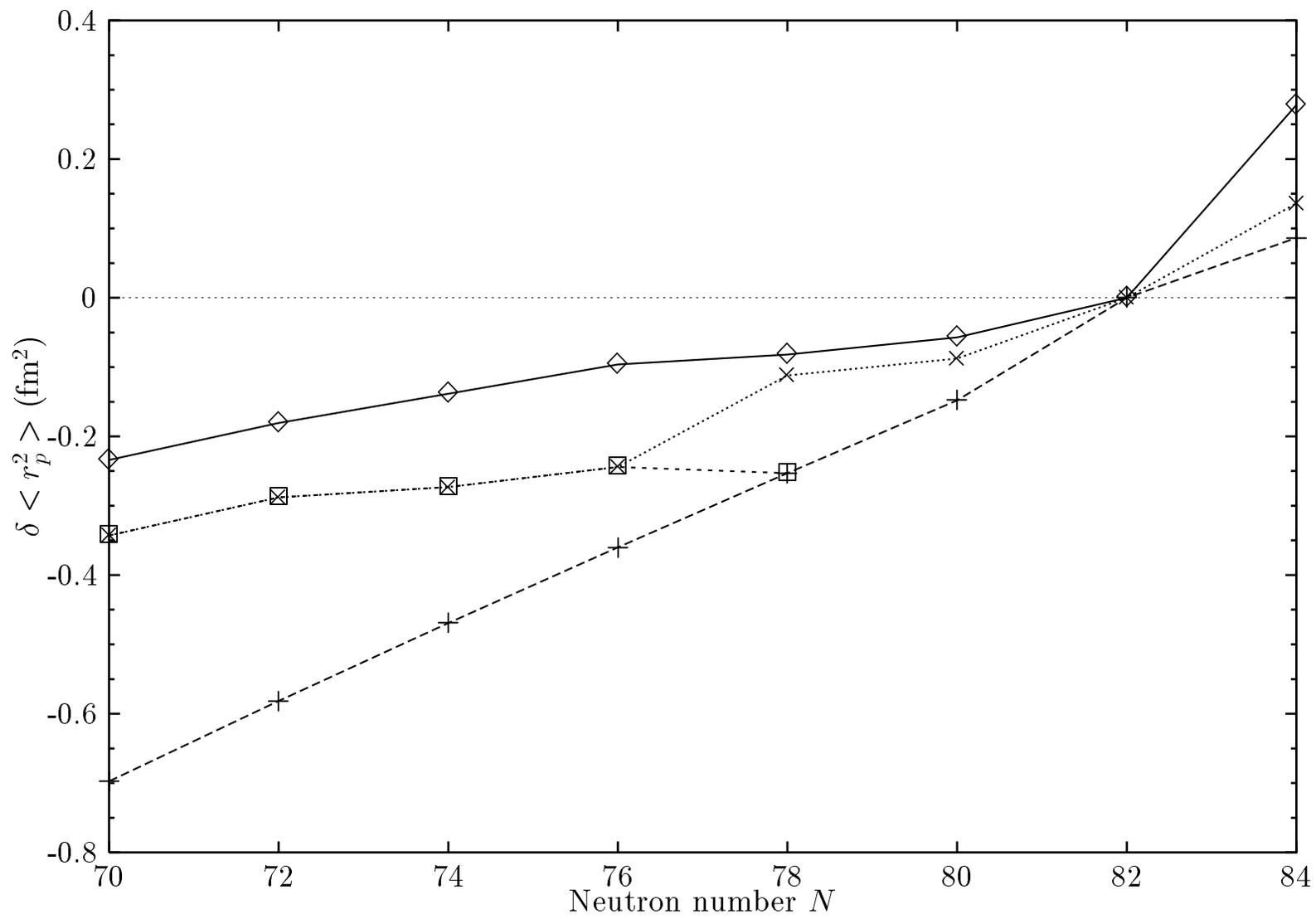
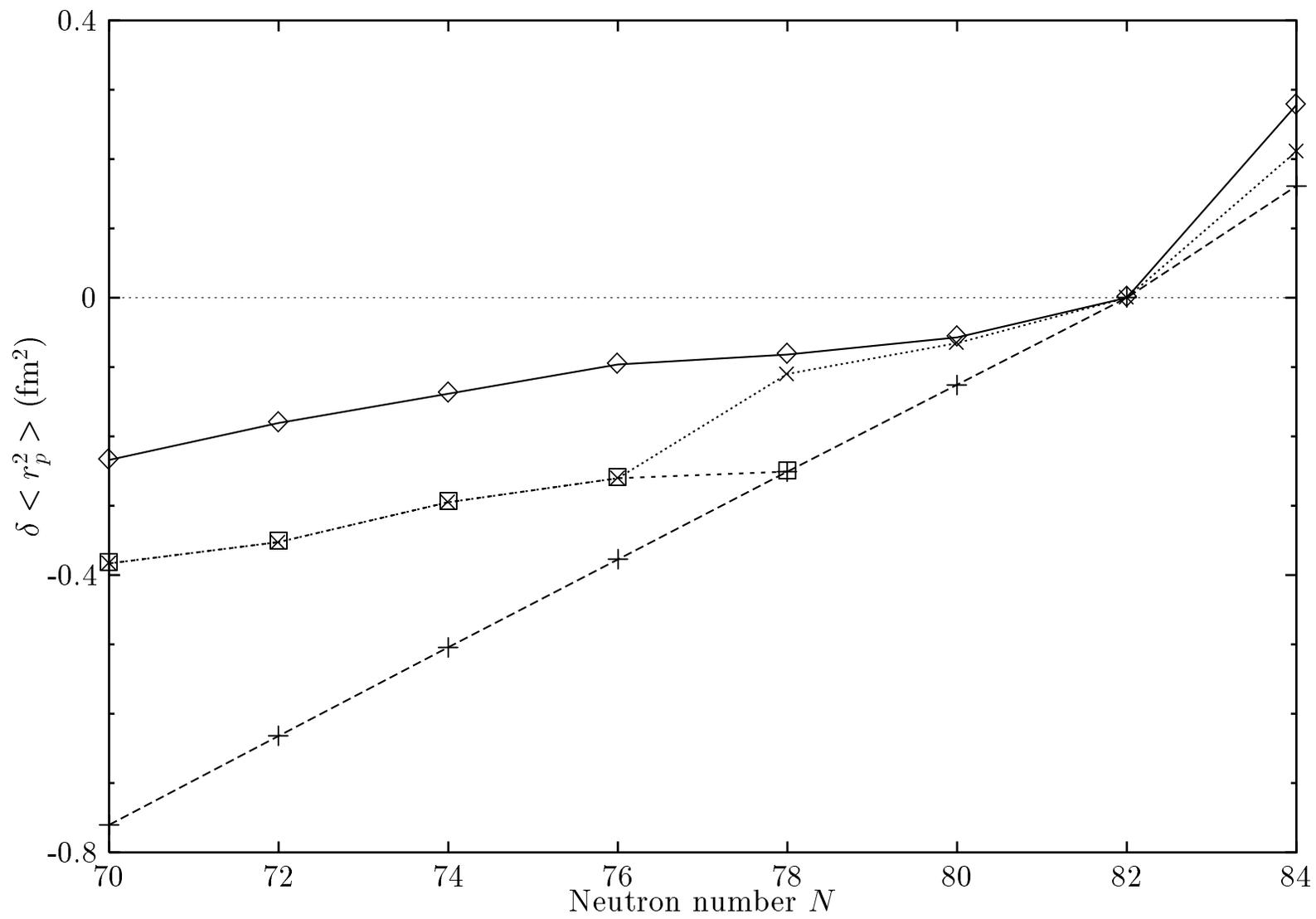
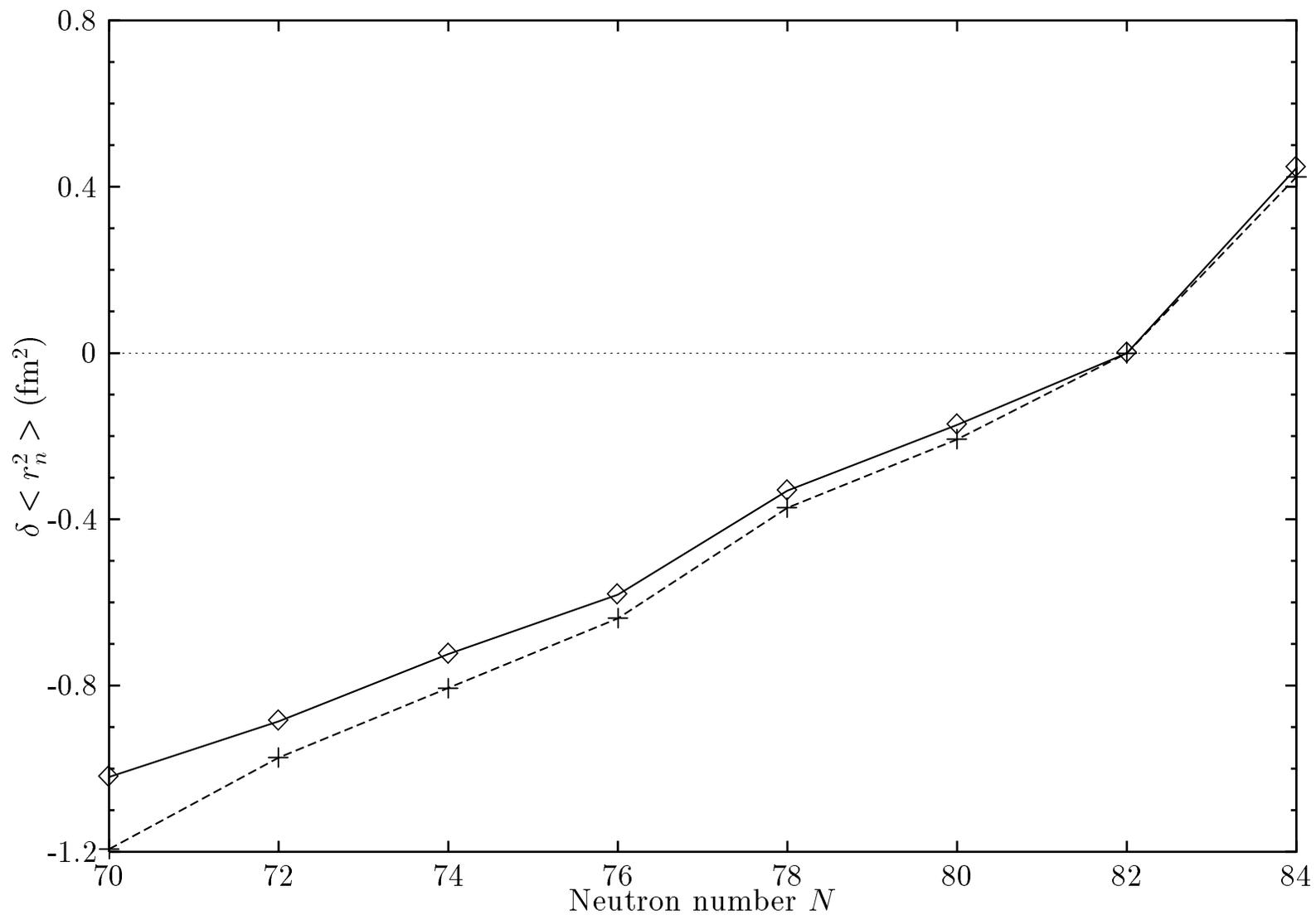
\end{document}